\documentclass[os]{copernicus}

\begin{document}

\title{Non-newtonian wind-driven flows in homogeneous semienclosed basins}


\Author[1,2]{Víctor J.}{Llorente}
\Author[2]{Enrique M.}{Padilla}
\Author[2]{Manuel}{Díez-Minguito}

\affil[1]{University of Zaragoza, C. de Pedro Cerbuna, 12, 50009 Zaragoza, Spain}
\affil[2]{Andalusian Institute for Earth System Research (IISTA), University of Granada, Avda. del Mediterráneo s/n, Edificio CEAMA, Granada E-18006, Spain}

\correspondence{Manuel Díez-Minguito (mdiezm@ugr.es)}

\runningtitle{Non-newtonian wind-driven flows}

\runningauthor{M.Díez-Minguito}

\received{}
\pubdiscuss{} 
\revised{}
\accepted{}
\published{}


\firstpage{1}

\maketitle

 \nolinenumbers

\begin{abstract}

Wind-driven flow in power-law viscous fluids in homogeneous semienclosed basins is analyzed. Analytical solutions for vertical current profiles for non-newtonian fluids with different power-law indexes are derived assuming a wind shear stress at the surface and zero net transport. Previous classical solutions for wind-driven flows in semienclosed basins are thus generalized. A bidirectional flow in the fluid column is obtained: downwind near the surface and a weaker compensating upwind flow near the bottom. Shear-thickening fluids exhibit larger maximum currents at the surface, although reducing the difference between the maximum current at the surface and the maximum current near the bottom, than newtonian and shear-thinning fluids. The larger the index $n$ of the power-law, the more acute the effects. For the same physical conditions (depth, wind shear stress, and flow consistency index), the opposite occurs for shear-thinning fluids. The lower the $n$ value the lower the current at the surface and the larger the differences between the upper and lower layer. 

\end{abstract}


\section{Introduction} 

The vertical shear of wind velocity provides locally a vertical transfer of horizontal momentum through turbulent stresses to water columns under gravity. Winds also can produce the movement of water as they drag surface waters downwind which, in turn, can set up a water level. This set-up, if stationary conditions hold in a (semi)enclosed basin, induces a gradient pressure force that drives an upwind flow which may flow underneath the surface flow. Classical vertical velocity profiles for environmental wind-driven flows in semienclosed basins, such as in certain types of estuaries, were derived by a number of authors in the past \citep{hansen1966gravitational,officer1976physical} and are widely used either as a reference or to gain deeper understanding on wind-driven flow processes \citep[e.g.][]{geyer1997influence,burchard2010quantifying,cheng2010upwelling,diez2020observations}. They considered a momentum balance with no inertia (low Rossby number) and no rotation in a laterally homogeneous basin (wind blowing along the basin) in which vertical frictional effects that overwhelm horizontal frictional effects are balanced by a gradient pressure force induced by wind.

High density particulate matter suspensions in water, such as mud layers, gravity currents and debris flows \citep[e.g.][]{winterwerp2002scaling,sheremetagu2008,di2006viscous}, quicksand \citep[e.g.][]{khaldoun2005liquefaction,kadau2009living}, masses of liquids different from water driven by planetary atmospheric dynamics, such as liquid methane lakes in Titan, a moon of Saturn \citep[e.g.][]{tokano2015wind,nimmo2016ocean}, are examples of non-newtonian fluids that can be driven by winds in environmental systems. Contrary to newtonian fluids, non-newtonian fluids exhibit a rheology in which apparent viscosity that varies with the shear rate. The relation between shear stress and shear rate is often conveniently described as a simple power law, i.e. Ostwald power-law model, which has been successfully applied to a variety of geophysical and industrial problems \citep[e.g.][]{bird1961transport}. In this manuscript we generalize classical solutions for wind-driven flows in homogeneous semienclosed basins for non-newtonian power-law fluids with arbitrary flow behavior index.

\section{Newtonian Wind-driven Flow} 

Classical solutions for the vertical velocity profiles for wind-driven flows in semienclosed basins are derived assuming a momentum balance with no inertia and no rotation in which vertical frictional effects dominate horizontal friction. The momentum equation states thus the balance between the gradient pressure force and vertical friction force as follows,
\begin{equation}\label{eq:momentum}
0=-g\rho \frac{\partial \eta}{\partial x} +\frac{\partial \tau}{\partial z}\,,
\end{equation}
with 
\begin{equation}\label{eq:tau}
\tau=\rho A\frac{\partial u}{\partial z}\,.
\end{equation}
Here $u=u(x,z)$ is the flow velocity, $\eta$ is the elevation of the free surface (set up), $\tau$ is the shear stress that corresponds with a classical newtonian fluid in which the shear stress depends linearly on the strain rate, $g$ is the gravitational acceleration, and $A$ is the vertical eddy viscosity. Assuming that the water-level slope $\partial \eta/\partial x$ is constant, the solution for $u$ can be obtained using Eq.~\ref{eq:tau} by integrating Eq.~\ref{eq:momentum} twice. The no slip at the bottom ($z=-H$) and wind stress boundary condition at the surface ($z=0$) read, respectively
\begin{align}\label{eq:bc1}
& u(x,-H)=0\,,\\ \label{eq:bc2}
& \frac{\partial u}{\partial z}=\frac{\tau_w}{\rho A}\,,
\end{align}
where $\rho$ is the liquid density and $\tau_w=\rho_a C_D w |w|$ is the wind stress with $w$ the wind velocity, $C_D$ a empirical drag coefficient, and $\rho_a$ the air density. Assuming a constant vertical eddy viscosity ($A=A_{0}$), the wind-induced vertical velocity profile $u(x,\zeta)$ reads
\begin{equation}\label{eq:uclasicaAzconst}
u(x,\zeta) = \frac{\tau_w H}{4\rho A_{0}}\left(3\zeta^2 + 4\zeta +1 \right)\,,
\end{equation}
where $\zeta=z/H$ and $H$ is the depth. The vertical coordinate $z$ is positive downwards and its origin, $z=0$, is set at the surface. For a given value of $A$, $\tau_w$ and $\partial \eta/\partial x$ are not independent. The set up is induced by the wind stress at the surface, i.e., they must be dynamically consistent. Assuming that in a given cross-section the net transport per unit basin width is zero, that is, 
\begin{equation}\label{eq:transport}
\int_{-1}^0 u(x,\zeta) \text{d}\zeta= 0\,,
\end{equation}
the relationship between the set-up and the wind forcing is 
\begin{equation}\label{eq:setupwind}
\frac{\partial \eta}{\partial x}=\frac{3}{2}\frac{\tau_w}{\rho g H}\,.
\end{equation}

\begin{figure*}[!htbp]
  \centering
  \includegraphics[width=\textwidth]{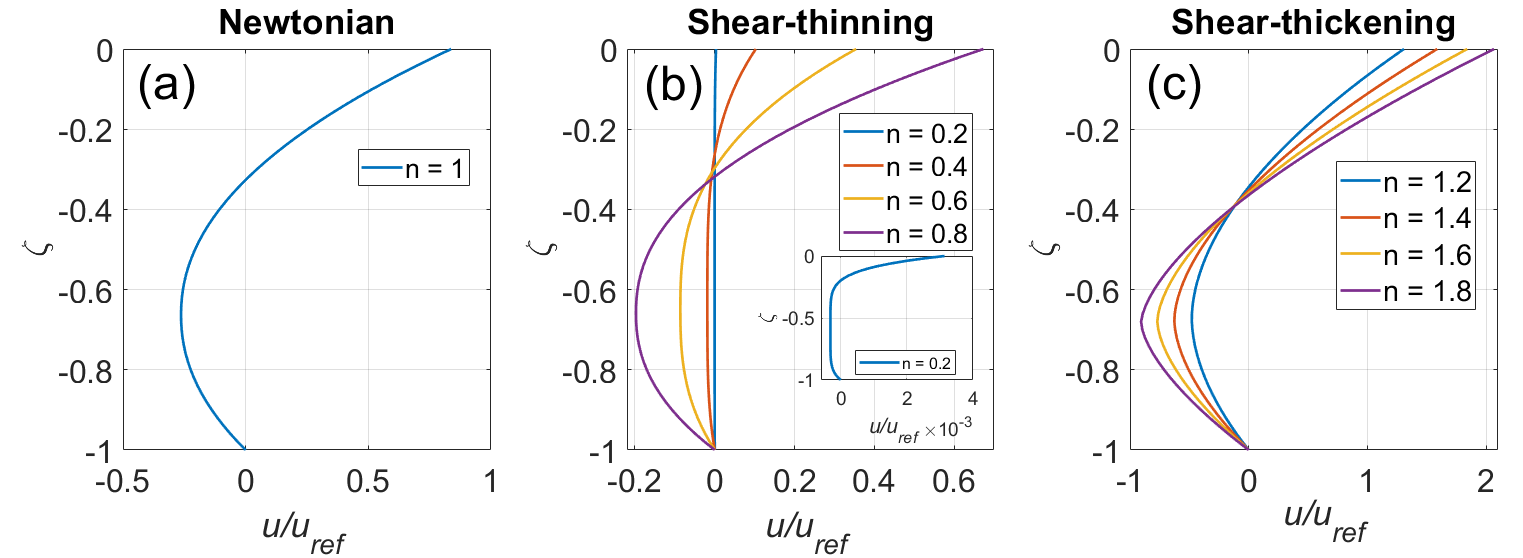}
  \caption{Scaled vertical current profiles for the classical newtonian ($n=1$, panel a), shear thinning ($n<1$, panel b), and shear thickening ($n>1$, panel c) fluids. Inset in panel b shows a zoom in the profile for $n=0.2$. Currents are scaled by $u_{ref}=\tau_w H/(4\rho A_{0})$ and the vertical coordinate $z$ is scaled by depth $H$.}
  \label{fig:perfilesu}
\end{figure*}

Figure~\ref{fig:perfilesu} (panel a) shows the scaled vertical current profile $u/u_{ref}$ with $u_{ref}=\tau_w H/(4\rho A_{0})$ as a function of the scaled depth $\zeta$ for the newtonian case with $n=1$ (Eq.~\ref{eq:uclasicaAzconst}). The flow near the surface is positive due to the positive wind shear stress that drags the free surface. The downwind flow near the surface is compensated by a upwind flow near the bottom. Although the thickness of the lower layer is larger than that of the upper layer, the profile verifies Eq.~\ref{eq:transport} and therefore the (depth-integrated) net transport is zero.

\section{Non-newtonian Wind-driven Flow}

Contrary to Newtonian fluids, non-newtonian fluids exhibit an apparent viscosity that varies with the shear rate $\partial u/\partial z$. The relation between shear stress and shear rate is often written as a power law which reads
\begin{equation}\label{eq:nonewton} 
\tau=\rho A\left|\frac{\partial u}{\partial z}\right|^{n-1}\frac{\partial u}{\partial z}\,. 
\end{equation}
Fluids that obey the Newton's law of viscosity have $n=1$. Fluids in which apparent viscosity decreases with higher shear rates, i.e., shear-thinning fluids, show $n<1$. Fluids in which viscosity increases with higher shear rates, i.e., shear-thickening fluids, show $n>1$.

The momentum balance equation when a generalized fluid is considered, i.e., using Eq.~\ref{eq:nonewton} in Eq.~\ref{eq:momentum}, reads
\begin{equation}\label{eq:momentum_non}
0=-g\frac{\partial \eta}{\partial x} +\frac{\partial }{\partial z}\left(A\left|\frac{\partial u}{\partial z}\right|^{n-1}\frac{\partial u}{\partial z}\right)\,.
\end{equation}
The same procedure as for newtonian fluids is followed in this section to obtain solutions of the vertical current profiles for non-newtonian fluids. Equation~\ref{eq:momentum_non} is integrated twice to obtain $u(x,\zeta)$. No-slip condition at the bottom (Eq.~\ref{eq:bc1}) and wind shear stress condition at the surface are considered. Notice that the latter extends Eq.~\ref{eq:bc2} to non-newtonian fluids. The boundary condition at the surface, by Eq.~\ref{eq:nonewton}, reads now $\left|\partial u/\partial z\right|^{n-1}\partial u\partial z=\tau_w/(\rho A)$ at $z=0$. Solutions for positive and negative values of $\partial u/\partial z$ need to be separated. Assuming both constant pressure gradient and eddy viscosity coefficient (i.e., the flow consistency index in a non-newtonian context), no-slip condition at the bottom and wind shear stress at the surface (Eq.~\ref{eq:bc1} and~\ref{eq:bc2}), the current profile is 
\begin{equation}\label{eq:profile_non_eta}
u(x,z)=\frac{n}{\alpha(n+1)}\left[\left|\alpha z+\beta\right|^{1+1/n}-\left|\alpha H+\beta\right|^{1+1/n}\right]\,,
\end{equation}
where $\alpha=gA^{-1}\partial \eta/\partial x$ and $\beta=\tau_w/(\rho A)$ and the change of sign in $\partial u/\partial z$ occurs at $z_c=-\beta/\alpha=-\tau_w/(\rho g \partial \eta/\partial x)$. Although the power-law model of Eq.~\ref{eq:nonewton} is recognized to show unphysical behavior for at very low and very high shear rates \citep{boger1977demonstration,tasmin2021non}, Eq.~\ref{eq:profile_non_eta} is valid for all $n>0$ values. 

The pressure gradient that is induced by the wind shear stress, i.e., the relationship between $\alpha$ and $\beta$, is obtained numerically. For a given wind shear stress $\tau_w$, the vertical current profile $u(x,z)$ is computed for different values of $\partial \eta/\partial x$. The optimum value of $\partial \eta/\partial x$ is that that makes $u(x,z)$ comply with Eq.~\ref{eq:transport}. The optimum value, and the corresponding vertical current profile that has zero transport, can be found with a simple Newton-Raphson iterative method. 

Figure~\ref{fig:perfilesu} (panels b and c) shows the scaled vertical current profile $u/u_{ref}$ (Eq.~\ref{eq:profile_non_eta}) as a function of the scaled vertical coordinate for the shear-thinning ($n<1$, panel b) and shear-thickening cases ($n>1$, panel c). As for the newtonian case ($n=1$, panel a), profiles in panels b and c show a positive flow near the surface due to the positive wind velocity. The zero transport conditions imply that near the bottom the flow should be in the opposite direction. For a constant value of $A=A_{0}$, the maximum positive (downwind) velocity always occurs at the surface, whereas the maximum negative (upwind) velocity always occurs at $z=-|\beta/\alpha|$ for all $n$. However, differently from the newtonian case, the slope of the profiles at the surface changes with $n$. 

Shear-thinning fluids ($n<1$, panel b), for the same values of depth, $H$, wind shear stress, $\tau_w$, and flow consistency index, $A_{0}$, show lower apparent viscosity values and lower current velocities overall than newtonian fluids. The lower the value of $n$ the lower the maximum positive and negative velocities, i.e., in both the upper downwind and lower upwind layers, respectively. The thickness of the upper layer decreases when $n$ decreases. Of all the vertical velocity profiles, the case for $n=0.2$ (inset panel b) shows the smallest thickness of the upper layer, and the greatest thickness of the lower layer. As $n$ decreases, the profile in the lower layer becomes more uniform. This is due to the fact that viscosity decreases with higher shear rates, i.e., shear stresses are damped by powers lower than 1. 

Wind-induced flows for different shear-thickening fluids are shown in panel c. For the same values of depth, $H$, wind shear stress, $\tau_w$, and flow consistency index, $A_{0}$, shear stress and maximum velocities are larger than for the newtonian case (and thus for $n<1$ cases). Similarly as for $n<1$, the thickness of the upper layer increases when $n$ increases. The larger the value of $n$ the larger the both maximum near-surface positive and maximum near-bottom negative currents. As $n$ increases, the profile in the lower layer becomes less uniform, thereby sharping at the maximum location at $-|\beta/\alpha|$. This is due to the fact that apparent viscosity is enhanced by higher shear rates.

\section{Conclusions} 

In this manuscript, the classical analytical solutions for wind-driven flows in homogeneous semienclosed basins were generalized. Non-newtonian power-law fluids with arbitrary flow behavior index subject to a wind shear stress at the surface and zero net transport were considered. The momentum balance assumes no inertia, no rotation, and that vertical frictional effects overwhelm horizontal friction, thereby existing a balance between the gradient pressure force and vertical friction force. Overall, the solutions derived allow to study the development of the bidirectional flow (i.e., downwind near the surface and a weaker compensating upwind flow near the bottom) as a function of the flow behavior index.

\begin{acknowledgements}
This research was funded by research projects EPICOS (Plan Andaluz de Investigación, Desarrollo e Innovación, PAIDI 2020. Ref. ProyExcel\_00375) and MUDDY (2ª Convocatoria de Ayudas a Proyectos de Investigación ``JÓVENES INVESTIGADORES CEIMAR2021''. Ref. CEIJ-009). VJL acknowledges the financial support from Ministerio de Universidades and Unión Europea-NextGenerationEU (Convocatoria de Recualificación Sistema Universitario Español-Convocatoria Complementaria ``Margarita Salas''). EMP acknowledges the financial support from the Plan Andaluz de Investigación, Desarrollo e Innovación (PAIDI 2020), funded by the European Union (Programa Operativo Fondo Social Europeo de Andalucía 2014-2020) and Junta de Andalucía.
\end{acknowledgements}

\end{document}